# Fundamentals of Order Dependencies

Jaroslaw Szlichta, Parke Godfrey, Jarek Gryz

York University, Computer Science & Engineering, Toronto, Canada
and
IBM Center for Advanced Studies, Toronto, Canada
{jszlicht, godfrey, jarek}@cse.yorku.ca

## ABSTRACT

Dependencies have played a significant role in database design for many years. They have also been shown to be useful in query optimization. In this paper, we discuss dependencies between lexicographically ordered sets of tuples. We introduce formally the concept of *order dependency* and present a set of axioms (inference rules) for them. We show how query rewrites based on these axioms can be used for query optimization. We present several interesting theorems that can be derived using the inference rules. We prove that functional dependencies are subsumed by order dependencies and that our set of axioms for order dependencies is sound and complete.

## 1. INTRODUCTION

Consider the following SQL query (in Example 1).

EXAMPLE 1.

```
select D.year, D.quarter, D.month,
    sum(S.sales) as total
from Dates D, Sales S
where D.date_id = S.date_id
    and D.year between 2001 and 2004
group by D.year, D.quarter, D.month
order by D.year, D.quarter, D.month
```

In the schema, `Dates` is a *dimension* table with a row per day, and `Sales` is a very large *fact* table recording all individual sales. Each has a surrogate-valued column `date_id`, which is the primary key for `Dates`. In the `Dates` dimension table, each row describes a given day with explicit columns as `year`, `quarter`, `month`, and `day` that describe the natural date values (and additional columns that qualify that day, such as whether it is a weekend day or holiday).

Assume we have a tree index for `Dates` on `year`, `month`, `day`. This index cannot help in a query plan, however, to accomplish the group-by because `quarter` intercedes. Of course, `quarter` is logically redundant here, as `month` (which follows it in the group-by) functionally determines `quarter`. (First quarter encompasses the months of January, February, and March, second quarter, the months of April, May, and June, and so forth.) The query's author could not leave `quarter` out of the group-by – even if he realizes it would be better to – because it is stated in the `select`. The query optimizer could, however, use an index scan to have the tuple stream in `year`, `month` order to accomplish the `group by` on `year`, `quarter`, `month`, *if* it recognizes that `year`, `month` and `year`, `quarter`, `month` offer the same partition. This is done by query optimizers today – given the functional dependency (FD) information that `month` → `quarter` is available to the optimizer – by rewrite.

For the query above, the rewrite might still not be applied, since the query specifies the answers to be ordered by `year`, *`quarter`*, `month`. The FD that `month` → `quarter` is *not* logically sufficient to eliminate `quarter` from the order-by, as it was to eliminate it from the group-by. Since a query plan must guarantee the order-by, it likely will include a sort operator for `year`, `quarter`, `month`, after all.

To see that the FD does not suffice to eliminate `quarter` from the order-by, imagine the values for `quarter` were the strings *first*, *second*, *third*, and *fourth*. Data would be lexicographically ordered as *first*, *fourth*, *second*, then *third*! Of course, we intend that values of `quarter` are, say, 1, 2, 3, and 4, so the data would order naturally as by date. It is unfortunate, then, that `quarter` is, in fact, redundant (in this query) in the order-by also, but that the optimizer does not have the means to eliminate it.

What is missing is the semantic information that `month` *orders* `quarter`, which is more than just that `month` *functionally determines* `quarter`. This states that as values *rise* from one tuple to another on `month`, they must *rise*, or stay the same, from the one tuple to the other on `quarter` (that is, the values do not descend from the one tuple to the other on `quarter`). These have been called *order dependencies* (ODs), in contrast to functional dependencies. Our objective is to bring reasoning about order dependencies into the query optimizer. A query plan for the query above could then eliminate `quarter` from *both* the order-by and the group-by clauses, and the index on `year`, `month`, `day` might then provide for an efficient plan with no need for a sort operator.

The notion of order dependencies can be greatly generalized, and the potential use of them in query optimization shown to be vast. The relationships between ordered sets have been explored in the past and several different notions of *order* have been considered. In this work, we consider just *lexicographical* ordering of tuples, as by the order-by operator in SQL, because this is the notion of order used in SQL and within query optimization for tuple streams.

The contribution of this paper is to present an *axiomatization* for order dependencies, analogous to Armstrong's axiomatization for FDs [1]. This provides a formal framework for reasoning about ODs. There are two reasons for one to pursue an axiomatization.

1. The axioms provide *insight* into how dependencies behave – and patterns for how dependencies logically follow from others – that are not easily evident reasoning from first principles.
2. A *sound* and *complete* axiomatization is the first necessary step to designing an efficient *inference procedure*.

Our axioms for ODs help us explore beneficial query rewrites. We show how they can be cast as a new type of integrity constraint to be used in query optimization. We derive theorems based on our axioms, which illustrate surprising inferences and equivalences over ODs, and which can provide for powerful query rewrites. While ODs for databases have been explored before, we present the first general axiomatization for them. We prove the *soundness* of the axioms. We demonstrate that Armstrong's axiomatization for FDs is subsumed within our axiomatization for ODs. (In this sense, ODs can be thought of as a generalization of FDs.) We then prove the *completeness* of the set of axioms. Working with ODs is more involved than with FDs because the order of the attributes matters. Thus, we must work with *lists* of attributes instead of with *sets*. This necessarily complicates our axioms – compared with Armstrong's axioms for FDs – and the proofs of our theorems.

**Outline.** In Section 2, we present (ODs) formally. We provide background, our notational conventions, and definitions for ODs (Section 2.1). We show from where ODs in databases naturally arise (Section 2.2). We demonstrate a number of effective ways ODs may be used in query optimization (Section 2.3). We discuss a query optimization technique with ODs that we have implemented as a prototype in IBM DB2 [18], and our ongoing work with these techniques. In Section 3, we introduce the axiomatization for ODs (Section 3.1), and we prove the soundness of the axioms (Section 3.2). We derive a collection of theorems using our axioms – which we use in the proof of completeness – which illustrate the utility of our axioms (Section 3.3). In Section 4, we prove the completeness of the axiomatization. We sketch our proof of completeness (Section 4.1). We demonstrate how FDs are *subsumed* within order dependencies (Section 4.2). With the requisite pieces in place, we present the formal proof of completeness of the axiomatization (Section 4.3). In Section 5, we discuss related work. In Section 6, we present plans for future work and make concluding remarks. This work, we feel, opens exciting venues for future work to develop a powerful new family of query optimization techniques in database systems.

## 2. ORDER DEPENDENCY

We first set out formal definitions for order dependencies that we need later in proofs. Next, we illustrate ODs in databases and how they arise. We then show the use-case scenarios for ODs for query optimization.

### 2.1 Formal Definitions

We adopt the notational conventions in Table 1. We consider a relation $\boldsymbol{R}$ with a schema set of attributes $\mathcal{U}$. Let **r** be an arbitrary *table instance over* $\boldsymbol{R}$; thus a *set* of tuples under $\boldsymbol{R}$'s schema with attributes $\mathcal{U}$. We limit table instances to *sets* in our definitions, to keep our definitions simpler and easier to follow. However, this could be changed to *multi-sets* easily, with no consequences to our axiomatization.

**Table 1. Notational conventions.**

*Relations*

- A capital letter in bold italics represents a *relation*: $\boldsymbol{R}$.
- A small letter in bold represent a *relational instance* (*a table*): **r**.
- We use capital letters to represent single *attributes*: A, B, C. *Tuples* are marked with small letters in italics: *s, t*.

*Sets*

- Calligraphic letters stand for *sets of attribute*: $\mathcal{X}, \mathcal{Y}, \mathcal{Z}$.
- We use proximity for *union* of sets: $\mathcal{XY}$ is shorthand for $\mathcal{X} \cup \mathcal{Y}$. Likewise, A$\mathcal{X}$ or $\mathcal{X}$A, where $\mathcal{X}$ is a set of attributes and A a single attribute, stands for $\mathcal{X} \cup \{\text{A}\}$.
- Also $t_{\mathcal{X}}$ denotes the projection of the tuple *t* on the attributes of $\mathcal{X}$, while $t_{\text{A}}$ is the shorthand for $t_{\{\text{A}\}}$.

*Lists*

- Bold letters stand for *lists* of attributes: **X**, **Y**, **Z**. Note list **X** could be the empty list, [].
- We use square brackets to denote a list: [A, B, C]. The notation [A | **T**] denotes that A is the *head* of the list, and **T** is the *tail* of the list, the remaining list with the first element removed.
- Proximity is used for concatenation of lists of attributes: **XY** is shorthand for $\mathbf{X} \circ \mathbf{Y}$. Likewise, A**X** and **X**A stands respectively for $[\text{A}] \circ \mathbf{X}$ and $\mathbf{X} \circ [\text{A}]$, where **X** is list of attributes and A is a single attribute. AB denotes [A, B].
- $\mathbf{X}'$ denotes some other permutation of elements of list **X**.

DEFINITION 1. (operator $\preccurlyeq$) Let **X** be a list of attributes, *s* and *t* be two tuples in relation instance **r**. Operator $\preccurlyeq$ is defined as follows:

$s_{\mathbf{X}} \preccurlyeq t_{\mathbf{X}}$ where $\mathbf{X} = [\text{A} \mid \mathbf{T}]$
 if $(s_{\text{A}} < t_{\text{A}})$
 or if $((s_{\text{A}} = t_{\text{A}})$ and $(\mathbf{T} = []$ or $s_{\mathbf{T}} \preccurlyeq t_{\mathbf{T}}))$

In this paper, we assume *ascending* (`asc`) order in the lexicographical ordering. (This is `SQL`'s default.) We do not consider descending (`desc`) orders, mixing of `asc` and `desc` (e.g., `order by X desc, Y asc`) [19], or use of functions in the order directives (e.g., `order by -1*X asc, Y asc`).

DEFINITION *2.* (operator $\prec$) Let **X** be a list of attributes, *s* and *t* be two tuples in relation instance **r**. The operator $\prec$ is defined as follows: $s_{\mathbf{X}} \prec t_{\mathbf{X}}$ *iff* $s_{\mathbf{X}} \preccurlyeq t_{\mathbf{X}}$ and $t_{\mathbf{X}} \not\preccurlyeq s_{\mathbf{X}}$.

DEFINITION 3. ($s_{\mathbf{X}} = t_{\mathbf{X}}$) Let **X** be a list of attributes, *s* and *t* be two tuples in relation instance **r**, $s_{\mathbf{X}} = t_{\mathbf{X}}$ iff $s_{\mathbf{X}} \preccurlyeq t_{\mathbf{X}}$ and $t_{\mathbf{X}} \preccurlyeq s_{\mathbf{X}}$.

DEFINITION 4. (order dependency) Let **X** and **Y** be list of attributes. Call $\mathbf{X} \mapsto \mathbf{Y}$ an *order dependency* (OD) over the relation $\boldsymbol{R}$ if, for every pair of *admissible* tuples *s* and *t* in relation instance **r** over $\boldsymbol{R}$, $s_{\mathbf{X}} \preccurlyeq t_{\mathbf{X}}$ implies $s_{\mathbf{Y}} \preccurlyeq t_{\mathbf{Y}}$.

Whenever $\mathbf{X} \mapsto \mathbf{Y}$, we say that **X** *orders* **Y**. **X** and **Y** are *order equivalent* iff $\mathbf{X} \mapsto \mathbf{Y}$ and $\mathbf{Y} \mapsto \mathbf{X}$. We denote this by $\mathbf{X} \leftrightarrow \mathbf{Y}$.

| A | B | C | D | E | F |
|---|---|---|---|---|---|
| 3 | 2 | 0 | 4 | 7 | 9 |
| 3 | 2 | 1 | 3 | 8 | 9 |

**Figure 1. Relation instance r.**

EXAMPLE 2. Note that [A, B, C] $\mapsto$ [F, E, D] is consistent with **r**, but [A, B, C] $\mapsto$ [F, D, E] is falsified by **r** in Figure 1.

The OD $\mathbf{X} \mapsto \mathbf{Y}$ means that **Y**'s values are *monotonically non-decreasing* with respect to **X**'s values. Thus, if a list of tuples are ordered by **Y**, then they are also necessarily ordered by **X**, but not necessarily vice versa. That is to say, if one knows $\mathbf{X} \mapsto \mathbf{Y}$, then one knows that any ordering of the tuples of **r**, for any **r**, that satisfies `order by` **Y** also satisfies `order by` **X**.

There is a clear relationship between ODs and FDs. Any OD implies and FD (modulo lists and sets), but not vice versa.

LEMMA 1. (relationship between ODs and FDs). *For every instance* **r** *of relation* $\boldsymbol{R}$*, if OD* $\mathbf{X} \mapsto \mathbf{Y}$ *holds, then FD* $\mathcal{X} \rightarrow \mathcal{Y}$ *is true.*

PROOF. Let $s$, $t \in \mathbf{r}$, such that $s_{\mathcal{X}} = t_{\mathcal{X}}$. Therefore, $s_{\mathbf{X}} \preccurlyeq t_{\mathbf{X}}$ and $t_{\mathbf{X}} \preccurlyeq s_{\mathbf{X}}$. By the definition of OD $s_{\mathbf{Y}} \preccurlyeq t_{\mathbf{Y}}$ and $t_{\mathbf{Y}} \preccurlyeq s_{\mathbf{Y}}$, hence as $s_{\mathbf{X}} = t_{\mathbf{X}}$, $s_{\mathcal{Y}} = t_{\mathcal{Y}}$. □

DEFINITION 5. (order compatible) Two lists **X** and **Y** are order compatible, denoted as $\mathbf{X} \sim \mathbf{Y}$ *iff* $\mathbf{XY} \leftrightarrow \mathbf{YX}$.

EXAMPLE 3. Note that [A, B] ~ [F, C] is consistent with **r**, but [A, C] ~ [F, D] is falsified by **r** in Figure 1.

## 2.2 Order By

The concept of functional dependencies has come to have profound importance in databases, especially in schema design. While functional dependencies are a simple notion in some ways, reasoning over them is, somewhat surprisingly, not nearly as simple. To gain insight into how sets of FDs behave, and to simplify the reasoning process over them, Armstrong provided an axiomatization for them [1]. Beyond layout and indexes, FDs play additional important roles in query optimization. Knowledge about prescribed FDs on the schema are used in the query-rewrite phase of optimization potentially to eliminate predicates. They are used in the cost-based phase to do better cardinality estimation. They are used also to recognize partitioning equivalences of tuple streams within query plans.

We have introduced ODs in analogy to FDs: functional dependencies are to group-by as order dependencies are to order-by. On the one hand, order is *not* important in the pure relational model on the *logical* side of the fence. Relational instances are *sets* of tuples. (Implemented systems allow for *multi-sets* of tuples, but again, there is no notion of order.) A schema is a *set* of attributes. SQL concedes a single order-by clause to be appended to a query to order the result set, as a convenience, given that people often want to see the results sorted in a given way. (This said, there are many places where order is *semantically* meaningful. *Data stream* extensions to the relational model make order a part of the model. For other data models such as XML – and XQuery over it – order is an integral part of the model.)

On the other hand, order plays pivotal roles on the physical side, in the physical database and in query optimization. Data is often stored sorted by a clustered (tree) index's key. In a query plan, an operator that takes as input the output stream of another operator can benefit in cases when the stream is sorted in a particular way. Aggregation queries (group-by) can be evaluated *on-the-fly* if the stream is ordered already in a way compatible with the requested `group-by` partition, rather than needing to do a partitioning operation that could involve heavy I/O expense.

Given $\mathbf{X} \mapsto \mathbf{Y}$, if one has an SQL query with `order by` **Y**, one can rewrite the query with `order by` **X** instead, and meet the intent of the original query. However, the rewritten query is *not* semantically equivalent the original (unless $\mathbf{X} \leftrightarrow \mathbf{Y}$)! One could not legally rewrite the query with `order by` **X** with `order by` **Y** instead. Strengthening the order-by conditions is permitted, but weakening them is not. (This is true, too inside query plans for ordered tuple streams.)

One does not need order *equivalences* then to accomplish useful query rewrites. Directional order dependencies (e.g., $\mathbf{X} \mapsto \mathbf{Y}$, but not $\mathbf{Y} \mapsto \mathbf{X}$) suffice. This makes ODs that much more versatile for rewrites. Notice this differs from the use of FDs for query rewrites, for instance, to simplify group-by's. To replace `year`, `quarter`, `month` by `year`, `month` in the group-by for the query in the example in Section 1, one should know the two are functionally *equivalent*. One could not replace it by `year`, `month`, `day`, for example, even though `{year, month, day}` → `{year, quarter, month}`.

Within query plans, group-by (partitions) can be accomplished either by a partition operation (such as by use of a hash index), or by the use of an ordered tuple stream (as provided by a tree-index scan or by a sort operation). When rewriting the partition criteria, if a partition operation is employed, the criteria must be equivalent. However, when an ordering operation is employed instead, then one has the same flexibility as noted for OD dependencies. Strengthening the criteria suffices. For instance, sorting by `year`, `month`, `day` would suffice to accomplish the group-by on `year`, `quarter`, `month`. (Group divisions can be found *on-the-fly* in the stream.)

An OD can be declared as an *integrity constraint* to prescribe which instances are admissible. (We have introduced this new type of constraint in a prototype branch of IBM DB2. See Section 2.3.) One can reason over ODs on relations in a similar way one now reasons about FDs over relations. Some order dependencies are *trivially true* [20]. That is, they are (trivially) satisfied by *any* table instance. For example, consider $\mathbf{XY} \mapsto \mathbf{X}$. Others are not trivial. If one knows a collection of order dependencies, $\mathcal{M}$ – declared as integrity constraints over relation $\boldsymbol{R}$ – one might soundly infer additionally order dependencies that must be *true* for $\boldsymbol{R}$. For example, if $\mathbf{X} \mapsto \mathbf{Y}$ and $\mathbf{Y} \mapsto \mathbf{Z}$ are *true*, then $\mathbf{X} \mapsto \mathbf{Z}$ is *true* also. (That is, ODs are transitive.)

While order is not part of the relational model, per se, ordered value domains are of key importance for most databases, and most queries. Many types of ODs are apparent in the semantics of databases (even though these ODs are not declared explicitly). Perhaps the most important of these ordered domains in practice is *time*. *Time* and *date* (time at a coarser granularity) are richly supported in the SQL standards. The common benchmark TPC-DS has 99 queries. Of these, 85 involve date operators and predicates (and five involve time operators and predicates). This is common for data-warehouses. Even if we were just limited to ODs over the date/time domain, we could derive great benefits in query optimization.

Figure 2 represents possible ODs, in which the left-hand side of a dependency is *time* and the right-hand side is one of the paths through the diagram. Each node is an equivalent class of the list of attributes leading up to it, with respect to the starting point. Theorem 10 proves that any list appearing on the left side can be suffixed by attributes appearing along an equivalent path. This is shown in Example 4.

EXAMPLE 4. As [time] ↦ [date, hour] holds and [date] ↦ [year, month, day], it follows (from Theorem 10 below) that [time] ↦ [date, month, hour].

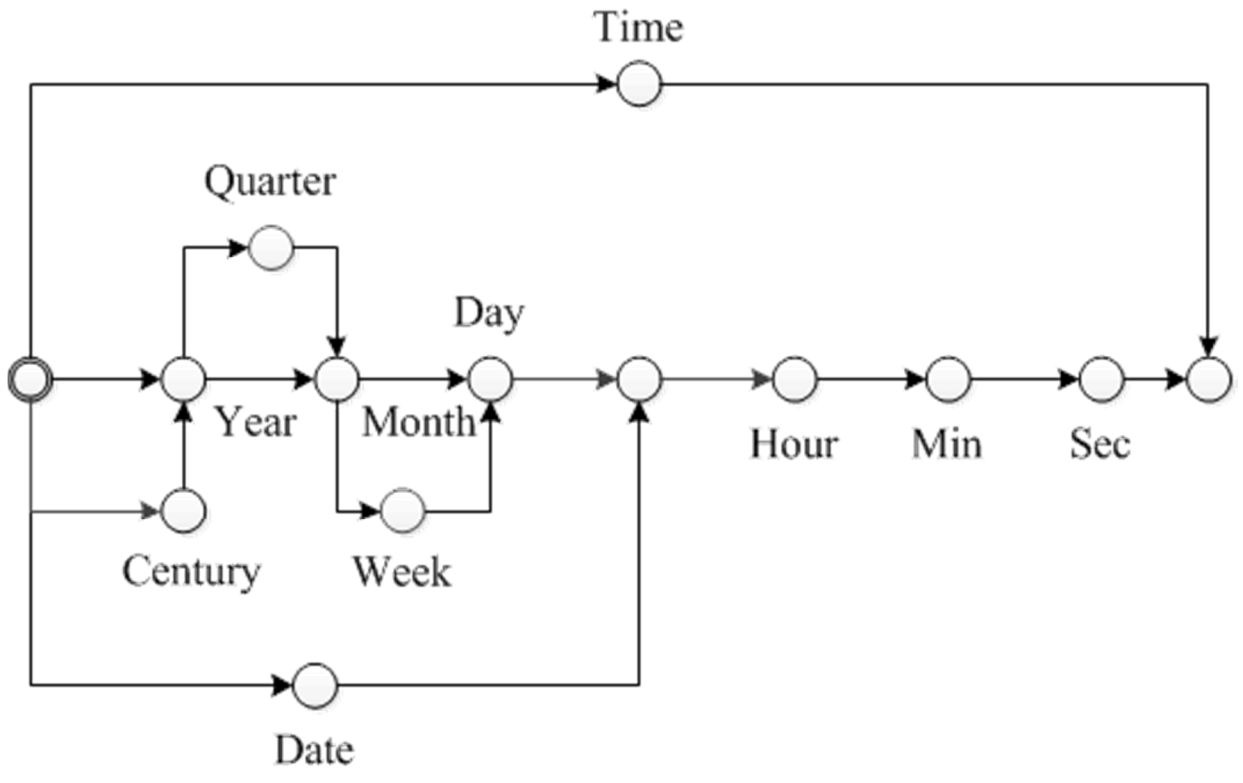

**Figure 2. Time diagram.**

Order dependencies are not just limited to the time domain, however. They arise naturally in many other domains from the real-world semantics associated with given data. All that is required is that the values of a column (or list of columns) are monotonically non-decreasing with respect to the values of another column (or list of columns). This property is fairly common when columns are functionally related.

EXAMPLE 5. Consider a table Taxes that includes columns for taxable income, tax bracket, and taxes on the income. The tax brackets are based on the level on income (and so rise with income level). Assume taxes go up with income. Then, [income] ↦ [bracket] and [income] ↦ [taxes]. It follows (from Theorem 2 below) that [income] ↦ [bracket, taxes].

Assume the table has a tree index on income. Given a query on the table with an order-by on bracket, taxes, with the OD above, it could be evaluated using the index on income.

Instead of being columns with explicit data, bracket and taxes could be derived by functions or case expressions – say, if Taxes were a view – or generated columns in the table. In these cases, it would be possible for the database system to derive the order-dependency constraints above automatically. In [12], it was shown how to derive such monotonicity "constraints" from generated columns via algebraic expressions (in IBM DB2). Of course, one could prescribe the set of order dependencies as check constraints directly to benefit by this technique.

Such monotonic dependencies can be derived from built-in SQL functions, from user-defined functions (to some degree), and from case expressions. The SQL function Year, for example, extracts the year component of a datestamp. Thus, given a datestamp column when, [when] ↦ [Year(when)].

## 2.3 Optimization

In [17], the authors expounded on the important role of *order* in query optimization. They demonstrated numerous examples of how better *reasoning* over *interesting orders* in the query optimizer could lead to significantly better performing query plans. They introduced query rewrites in IBM DB2 that could replace one labeled interesting order by another, when it is known the two order in the same way (that is, are *order equivalent*, as we have defined it).

They showed how these rewrites could allow the optimizer to consider additional query plans that process *join*, *order-by*, *group-by*, and *distinct* operators more efficiently. By recognizing that a tuple stream ordered with respect to some criteria is equivalently ordered with respect to other criteria, a sort on input can be removed for a sort-merge join. Order-by and group-by operators can be satisfied with no need for a sorting or partitioning operation more often, as with our Example 1. Likewise, as the distinct operator is exchangeable with group-by, the need for a sorting or partitioning operation to satisfy distinct can be lessened.

Our work builds upon this work. Their rewrites rely on functional dependency information available to the optimizer, but do not exploit any *order dependency* semantics, as defined by us. Our work permits a greater range of rewrites. For example, they could reduce an order-by year, month, quarter to an order-by year, month, based upon the FD {month} → {quarter}. (Likewise, they could reduce the equivalent group-by.) However, they could not reduce the order-by year, quarter, month to year, month, as we did in Example 1, since their techniques do not employ the idea of ODs. (It is Theorem 8 below, called *Left Eliminate*, which follows from our axiomatization, which justifies this rewrite.)

In [17], they introduced a rewrite algorithm for order-by called *Reduce Order*. It sweeps the order-by attribute list from right to left, seeking to eliminate attributes. Each iteration through the list, the prefix *set* with respect to the *current* attribute – that is, the set of attributes to the left of the current – is checked to see whether it *functionally determines* the current attribute. If so, the attribute is dropped from the list.

We can augment that algorithm – call it *Reduce Order** – to do an additional step. Each iteration through the list, it can additionally be checked whether any postfix *list* with respect to the current attribute – that is, a list of attributes to the right of the current – *orders* the current attribute. If so, the attribute is dropped from the list. Given the OD [month] ↦ [quarter], both order-by year, month, quarter and year, quarter, month would be reduced to year, month.

Order dependencies are in terms of lists of attributes, not sets as for functional dependencies. This makes matching in rewrites using ODs more complex generally, but also increases the possibilities for matches. Consider D ↦ B. Then ABD could be reduced to AD. However, ABCD cannot be! The attribute C intervening between the B and D invalidates the rewrite. For the rewrite by Theorem 8 to apply, the list on the right-hand side of the OD must precede directly the list on the left-hand side. If we knew D ↦ BC, then ABCD could be reduced to AD.

A major part of our continued work with order dependencies is to develop a number of efficient rewrite rules for the query optimizer, as they did in [17], to exploit ODs effectively. Our OD axiomatization provides us the means now to pursue this. The axioms and related theorems as in Section 3.3 provide us with insight into the types of rewrites that are possible. In [18], we developed query rewrites in a prototype branch of the IBM DB2 9.7 codebase that demonstrates the effectiveness of rewrites using order equivalences. In data-warehouses, date is often represented explicitly as a dimension table of its own, with the primary key of the date table made as a surrogate key [11]. While this design can have compelling advantages, the surrogate key can cause problems for efficiently evaluating queries.

A majority of queries in a data warehouse are over the fact table. A query often uses natural date values in predicates. However, date in the fact table is recorded by the surrogate key. This necessitates potentially a quite expensive join between the fact table and the date dimension table when the query is evaluated. There is an additional problem when a fact table has been partitioned by date in order to accommodate a very large table (e.g., in distributed systems). Since the date range (surrogate values) over the fact table cannot be determined from the query (natural values), all partitions of the fact table must be scanned. We optimize such queries involving dates by removing this join, and choosing just the relevant partitions of the fact table when the table is distributed.

A number of queries in the TPC-DS benchmark have this condition. Fortunately, we have a guarantee (an OD) that the surrogate (date) keys in the date dimension table are ordered in the same way as natural date values in the dimension table. Thus, the query plan can make two probes into the dimension table to calculate the range of the surrogate keys from the fact table. These two probes into the date table find `mindate` and `maxdate` surrogate key values. These two surrogate key values replace the range predicate, which allows the index on the date column in the fact table to be used.

The details of when and how this rewrite can be performed in a general way are provided in [18]. We built a prototype implementing such rewrites in IBM DB2 V.9.7 and performed experiments over TPC-DS to demonstrate the efficiency of the approach. Thirteen of TPC-DS`s queries matched the conditions for this rewrite. Every one of these thirteen benefited, with an average performance gain of 48%. Since this work reported in [18], we have continued work on the prototype. We have added a new type of check constraint which expresses an OD. We have implemented more OD rewrite rules which now rewrite eighteen of TPC-DS's queries with performance gain. Consider our technique from [18] combined with an OD rewrite of the order-by for our query in Example 1. If we have the OD that `[date_id]` ↦ `[year, month]`, the order-by and group-by operators in a query plan could be accomplished by an index scan over the index for `Sales`, the fact table, on `date_id`, then joining the results against the dimension table `Dates`.

# 3. AXIOMATIZATION

A key concern in dependency theory is developing the algorithms for testing logical implication. Developing inference rules is an approach to show logical implication between dependencies.

## 3.1 Axioms

DEFINITION 6. (A proof of OD θ from $\mathcal{M}$) Let $\mathcal{M}$ be a set of prescribed ODs. A proof of OD $\theta$ from $\mathcal{M}$ with the set of inference rules $\mathcal{I}$ is a sequence $\theta = \theta_1, \dots, \theta_n$ $(n \geq 1)$ such that for $k \in [1, n]$ either $\theta_k \in \mathcal{M}$, or there exists a substitution for some rule $\theta \in \mathcal{I}$ , such that $\theta_k$ is consequence of $\varphi$, and such that for each order dependency in the predecessor of $\theta$ the corresponding order dependency is in the set $\{\theta_i \mid 1 \leq i < k\}$.

The OD $\theta$ is provable from $\mathcal{M}$ using axioms $\mathcal{I}$ (relative to set of attributes *U*), denoted $\mathcal{M} \vdash_{\mathcal{I}} \theta$, if there is a proof of $\theta$ from $\mathcal{M}$ using $\mathcal{I}$. We now introduce axioms (inference rules) for ODs.

DEFINITION 7. (OD axioms) The inference rules for ODs are as follows.

**OD1:** *Reflexivity*

$$\mathbf{XY} \mapsto \mathbf{X}$$

**OD2:** *Prefix*

$$\frac{\mathbf{X} \mapsto \mathbf{Y}}{\mathbf{ZX} \mapsto \mathbf{ZY}}$$

**OD3:** *Normalization*

$$\mathbf{WXYXV} \leftrightarrow \mathbf{WXYV}$$

**OD4:** *Transitivity*

$$\frac{\mathbf{X} \mapsto \mathbf{Y} \qquad \mathbf{Y} \mapsto \mathbf{Z}}{\mathbf{X} \mapsto \mathbf{Z}}$$

**OD5:** *Suffix*

$$\frac{\mathbf{X} \mapsto \mathbf{Y}}{\mathbf{X} \leftrightarrow \mathbf{YX}}$$

**OD6:** *Chain*

$$\frac{\mathbf{X} \sim \mathbf{Y_1} \qquad \forall_{i \in [1, n-1]} \mathbf{Y}_i \sim \mathbf{Y}_{i+1} \qquad \mathbf{Y}_n \sim \mathbf{Z} \qquad \forall_{i \in [1, n]} \mathbf{Y}_i\mathbf{X} \sim \mathbf{Y}_i\mathbf{Z}}{\mathbf{X} \sim \mathbf{Z}}$$

Two of our axioms generate trivial dependencies [20]: Reflexivity and Normalization. We define the closure of the set of OD $\mathcal{M}$, denoted $\mathcal{M}^+$, to be the set of ODs that are logically implied by $\mathcal{M}$.

DEFINITION 8. (closure of $\mathcal{M}$ using $\mathcal{I}$). Let $\mathcal{I}$ = {OD1–OD6}, then $\mathcal{M}^+ = \{\mathbf{X} \mapsto \mathbf{Y} \mid \mathcal{M} \vdash_{\mathcal{I}} \mathbf{X} \mapsto \mathbf{Y}\}$.

DEFINITION 9. (equivalents sets of OD). Let $\mathcal{M}$ and $\mathcal{M}'$ be sets of ODs. We say that $\mathcal{M}$ and $\mathcal{M}'$ are equivalent *iff* $\{\mathbf{X} \mapsto \mathbf{Y} \mid \mathcal{M} \vDash \mathbf{X} \mapsto \mathbf{Y}\} = \{\mathbf{X} \mapsto \mathbf{Y} \mid \mathcal{M}' \vDash \mathbf{X} \mapsto \mathbf{Y}\}$.

## 3.2 Soundness

In this subsection, we address the problem of showing that our OD axioms are sound. This is to say, they lead only to true conclusions.

DEFINITION 10 (soundness of OD axioms) Let $\mathcal{I}$ be a set of inference rules {OD1–OD6}. Then $\mathcal{I}$ is sound for logical implication of ODs if $\mathbf{X} \mapsto \mathbf{Y}$ is deduced from $\mathcal{M}$ ($\mathcal{M} \vdash_{\mathcal{I}} \mathbf{X} \mapsto \mathbf{Y}$) using axioms $\mathcal{I}$, then $\mathbf{X} \mapsto \mathbf{Y}$ is true in any relation in which the dependencies of $\mathcal{M}$ are true $\mathcal{M} \vDash \mathbf{X} \mapsto \mathbf{Y}$.

Let **r** be a relation over ***R***. The following Lemmas are true.

LEMMA 2. (soundness of Reflexivity) *Reflexivity is sound.*

PROOF. Let *s, t* ∈ **r**, such that $s_{\mathbf{XY}} \preccurlyeq t_{\mathbf{XY}}$. From the recursiveness of Definition 1 of operator $\preccurlyeq$ it follows that (1) $s_{\mathbf{X}} = t_{\mathbf{X}}$ and $s_{\mathbf{Y}} \preccurlyeq t_{\mathbf{Y}}$ or (2) $s_{\mathbf{X}} \prec t_{\mathbf{X}}$. (1) and (2) imply that $s_{\mathbf{X}} \preccurlyeq t_{\mathbf{X}}$, therefore $\forall \mathbf{r} \,.\, \mathbf{XY} \mapsto \mathbf{X}$. □

LEMMA 3. (soundness of Prefix) *Prefix is sound.*

PROOF. Let *s, t* ∈ **r**, such that $s_{\mathbf{ZX}} \preccurlyeq t_{\mathbf{ZX}}$. This implies (1) $s_{\mathbf{Z}} \prec t_{\mathbf{Z}}$ or (2) $s_{\mathbf{Z}} = t_{\mathbf{Z}}$ and $s_{\mathbf{X}} \preccurlyeq t_{\mathbf{X}}$. For (1) $s_{\mathbf{ZY}} \preccurlyeq t_{\mathbf{ZY}}$ holds as $s_{\mathbf{Z}} \prec t_{\mathbf{Z}}$. In the second scenario (2), $s_{\mathbf{X}} \preccurlyeq t_{\mathbf{X}}$ implies $s_{\mathbf{Y}} \preccurlyeq t_{\mathbf{Y}}$ ($\mathbf{X} \mapsto \mathbf{Y}$ is given). Hence, as $s_{\mathbf{Z}} = t_{\mathbf{Z}}$ it is true that $s_{\mathbf{ZY}} \preccurlyeq t_{\mathbf{ZY}}$. $\forall \mathbf{r} \,.\, \mathbf{X} \mapsto \mathbf{Y}$ implies $\mathbf{ZX} \mapsto \mathbf{Z}$. □

LEMMA 4. (soundness of Normalization) *Normalization is sound.*

PROOF. **(IF)** Let *s, t* ∈ **r**, such that $s_{\mathbf{WXYV}} \preccurlyeq t_{\mathbf{WXYV}}$. This implies that: (1) $s_{\mathbf{WXY}} = t_{\mathbf{WXY}}$ and $s_{\mathbf{V}} \preccurlyeq t_{\mathbf{V}}$ or (2) $s_{\mathbf{WXY}} \prec t_{\mathbf{WXY}}$. In (1) $s_{\mathbf{X}} = t_{\mathbf{X}}$ as $s_{\mathbf{WXY}} = t_{\mathbf{WXY}}$. Therefore we can suffix **WXY** by list **X** and $s_{\mathbf{WXYX}} = t_{\mathbf{WXYX}}$ holds. Hence, $s_{\mathbf{WXYXV}} \preccurlyeq t_{\mathbf{WXYXV}}$ as we know that $s_{\mathbf{V}} \preccurlyeq t_{\mathbf{V}}$. Scenario (2), as $s_{\mathbf{WXY}} \prec t_{\mathbf{WXY}}$ implies that we can suffix list **WXY** by **XV** and $s_{\mathbf{WXYXV}} \preccurlyeq t_{\mathbf{WXYXV}}$ holds.

**(ONLY IF)** Let *s, t* ∈ **r**, such that $s_{\mathbf{WXYXV}} \preccurlyeq t_{\mathbf{WXYXV}}$. This implies that: (1) $s_{\mathbf{WXY}} = t_{\mathbf{WXY}}$ and $s_{\mathbf{XV}} \preccurlyeq t_{\mathbf{XV}}$ or (2) $s_{\mathbf{WXY}} \prec t_{\mathbf{WXY}}$. In (1) $s_{\mathbf{X}} = t_{\mathbf{X}}$ as $s_{\mathbf{WXY}} = t_{\mathbf{WXY}}$. Hence, $s_{\mathbf{V}} \preccurlyeq t_{\mathbf{V}}$ as we know that $s_{\mathbf{XV}} \preccurlyeq t_{\mathbf{XV}}$. Therefore, $s_{\mathbf{WXYV}} \preccurlyeq t_{\mathbf{WXYV}}$. Scenario (2), as $s_{\mathbf{WXY}} \prec t_{\mathbf{WXY}}$ implies that we can suffix list **WXY** by **V** and $s_{\mathbf{WXYV}} \preccurlyeq t_{\mathbf{WXYV}}$ holds. $\forall \mathbf{r}\,.\,\mathbf{WXYXV} \leftrightarrow \mathbf{WXYV}$. □

LEMMA 5. **(**soundness of Transitivity**)** *Transitivity is sound.*

PROOF. Let *s, t* ∈ **r**, such that $s_{\mathbf{X}} \preccurlyeq t_{\mathbf{X}}$. By $\mathbf{X} \mapsto \mathbf{Y}$ which is given $s_{\mathbf{Y}} \preccurlyeq t_{\mathbf{Y}}$ which implies $s_{\mathbf{Z}} \preccurlyeq t_{\mathbf{Z}}$ and it ends the proof. $\forall \mathbf{r}\,.\,\mathbf{X} \mapsto \mathbf{Y} \wedge \mathbf{Y} \mapsto \mathbf{Z}$ implies $\mathbf{X} \mapsto \mathbf{Z}$. □

LEMMA 6. **(**Soundness of Suffix**)** *Suffix is sound.*

PROOF. **(IF)** Let *s, t* ∈ **r**, such that $s_{\mathbf{X}} \preccurlyeq t_{\mathbf{X}}$. Therefore $s_{\mathbf{Y}} \preccurlyeq t_{\mathbf{Y}}$ as $\mathbf{X} \mapsto \mathbf{Y}$ is given, which implies that $s_{\mathbf{YX}} \preccurlyeq t_{\mathbf{YX}}$ ($\mathbf{X} \mapsto \mathbf{YX}$).

**(ONLY IF)** Let *s, t* ∈ **r**, such that $s_{\mathbf{YX}} \preccurlyeq t_{\mathbf{YX}}$. Therefore (1) $s_{\mathbf{Y}} = t_{\mathbf{Y}}$ and $s_{\mathbf{X}} \preccurlyeq t_{\mathbf{X}}$ or (2) $s_{\mathbf{Y}} \prec t_{\mathbf{Y}}$ is true. Scenario (1) directly implies that $s_{\mathbf{X}} \preccurlyeq t_{\mathbf{X}}$ ($\mathbf{YX} \mapsto \mathbf{X}$). Scenario (2) where $s_{\mathbf{Y}} \prec t_{\mathbf{Y}}$ implies that $s_{\mathbf{X}} \prec t_{\mathbf{X}}$. This is because $s_{\mathbf{X}} \nprec t_{\mathbf{X}}$ implies $t_{\mathbf{X}} \prec s_{\mathbf{X}}$ which implies $t_{\mathbf{Y}} \prec s_{\mathbf{Y}}$. Hence $s_{\mathbf{Y}} \nprec t_{\mathbf{Y}}$. This ends the proof as $s_{\mathbf{X}} \preccurlyeq t_{\mathbf{X}}$ ($\mathbf{YX} \mapsto \mathbf{X}$). $\forall \mathbf{r}\,.\,\mathbf{X} \leftrightarrow \mathbf{YX}$ □

LEMMA 7. (Soundness of Chain) *Chain is sound.*

PROOF. Without loss of generality, assume that the lists in the axiom are single attributes. Let $\mathbf{X} = \mathrm{A}$, $\mathbf{Y}_1 = \mathrm{B}_1, \ldots, \mathbf{Y}_n = \mathrm{B}_n$ and $\mathbf{Z} = \mathrm{C}$ This simplification makes it easier to extend the rule to lists. The proof is by contradiction. Assume that A and C are order incompatible. Then there are two tuples for which there is a swap (The notion of swap is formalized in Definition 14) of the values between A and C. Also the two tuples disagree on attribute $\mathrm{B}_i$ for all *i*. Otherwise condition number 4 would not be true. As A ~ $\mathrm{B}_1$, the values for $\mathrm{B}_1$ follow A, so does the rest of attributes $\mathrm{B}_i$ because of the condition (2). This means the two rows look in the following way:

| A | $\mathrm{B}_1$ | $\mathrm{B}_2$ | … | $\mathrm{B}_n$ | C |
|---|---|---|---|---|---|
| 0 | 0 | 0 | 0 | 0 | 1 |
| 1 | 1 | 1 | 1 | 1 | 0 |

**Figure 3.** A **order incompatible with** C**.**

But then $\mathrm{B}_n$ is order incompatible with C, which we assumed not to be the case. We conclude with contradiction. □

THEOREM 1. (soundness). *OD1-OD6 axioms are sound for logical implication of ODs.*

PROOF. In order to prove the soundness of $\mathcal{I}$ we have to prove that each of the rules is sound. This is Lemma 2 – Lemma 7. □

## 3.3 Theorems

We introduce additional inference rules as they will be used throughout the paper.

THEOREM 2 (Union)

1 $\mathbf{X} \mapsto \mathbf{Y}$
2 $\mathbf{X} \mapsto \mathbf{Z}$
$\overline{\mathbf{X} \mapsto \mathbf{YZ}}$

PROOF.

| | | |
|---|---|---|
| 3 | $\mathbf{YX} \mapsto \mathbf{YZ}$ | [Pref(2)] |
| 4 | $\mathbf{X} \mapsto \mathbf{YX}$ | [Suf(1)] |
| | $\mathbf{X} \mapsto \mathbf{YZ}$ | [Tran(3, 4)] □ |

THEOREM 3. (Augmentation)

1 $\mathbf{X} \mapsto \mathbf{Y}$
$\overline{\mathbf{XZ} \mapsto \mathbf{Y}}$

PROOF.

| | | |
|---|---|---|
| 2 | $\mathbf{XZ} \mapsto \mathbf{X}$ | [Ref] |
| | $\mathbf{XZ} \mapsto \mathbf{Y}$ | [Tran(1,2)] □ |

THEOREM 4. (Shift)

1 $\mathbf{W} \leftrightarrow \mathbf{V}$
2 $\mathbf{X} \mapsto \mathbf{Y}$
$\overline{\mathbf{WX} \mapsto \mathbf{VY}}$

PROOF.

| | | |
|---|---|---|
| 3 | $\mathbf{VX} \mapsto \mathbf{W}$ | [Aug(1)] |
| 4 | $\mathbf{VVX} \mapsto \mathbf{VW}$ | [Pref(3)] |
| 5 | $\mathbf{VVX} \leftrightarrow \mathbf{VX}$ | [Norm] |
| 6 | $\mathbf{VX} \mapsto \mathbf{VW}$ | [Tran(4,5)] |
| 7 | $\mathbf{VX} \leftrightarrow \mathbf{VWVX}$ | [Suf(6)] |
| 8 | $\mathbf{VWX} \leftrightarrow \mathbf{VWVX}$ | [Norm] |
| 9 | $\mathbf{VX} \leftrightarrow \mathbf{VWX}$ | [Tran(7,8)] |
| 10 | $\mathbf{WX} \mapsto \mathbf{V}$ | [Aug(1) |
| 11 | $\mathbf{WX} \mapsto \mathbf{VWX}$ | [Suf(10)] |
| 12 | $\mathbf{WX} \mapsto \mathbf{VX}$ | [Tran(9,11)] |
| 13 | $\mathbf{VX} \mapsto \mathbf{VY}$ | [Pref(2)] |
| | $\mathbf{WX} \mapsto \mathbf{VY}$ | [Tra(12,13)] □ |

THEOREM 5. **(**Decomposition)

1 $\mathbf{X} \mapsto \mathbf{ZY}$
$\overline{\mathbf{X} \mapsto \mathbf{Z}}$

PROOF.

| | | |
|---|---|---|
| 2 | $\mathbf{ZY} \mapsto \mathbf{Z}$ | [Ref] |
| | $\mathbf{X} \mapsto \mathbf{Z}$ | [Tran(1,2)] □ |

The following theorem is helpful to prove the Eliminate, Left Eliminate and Drop.

THEOREM 6. (Replace)

1 $\mathbf{M} \leftrightarrow \mathbf{N}$
$\overline{\mathbf{XMZ} \leftrightarrow \mathbf{XNZ}}$

PROOF.

| | | |
|---|---|---|
| 2 | $\mathbf{Z} \mapsto \mathbf{Z}$ | [Ref] |
| 3 | $\mathbf{MZ} \mapsto \mathbf{NZ}$ | [Shift(1,2)] |
| 4 | $\mathbf{NZ} \mapsto \mathbf{MZ}$ | [Shift(1,2)] |
| 5 | $\mathbf{XMZ} \mapsto \mathbf{XNZ}$ | [Pref(3)] |
| 6 | $\mathbf{XNZ} \mapsto \mathbf{XMZ}$ | [Pref(4)] |
| | $\mathbf{XMZ} \leftrightarrow \mathbf{XNZ}$ | [Tran(5,6)] □ |

THEOREM 7. (Eliminate)

1 $\mathbf{X} \mapsto \mathbf{Y}$
$\overline{\mathbf{MXNYW} \leftrightarrow \mathbf{MXNW}}$

PROOF.

| | | |
|---|---|---|
| 2 | $\mathbf{X} \leftrightarrow \mathbf{YX}$ | [Suf(1)] |
| 3 | $\mathbf{XX} \mapsto \mathbf{XYX}$ | [Pref(2)] |
| 4 | $\mathbf{X} \leftrightarrow \mathbf{XX}$ | [Norm] |
| 5 | $\mathbf{XY} \leftrightarrow \mathbf{XYX}$ | [Norm |
| 6 | $\mathbf{X} \leftrightarrow \mathbf{XY}$ | [Tran(3-5)] |
| 7 | $\mathbf{MXYNYW} \leftrightarrow \mathbf{MXNYW}$ | [Rep(6)] |
| 8 | $\mathbf{MXYNYW} \leftrightarrow \mathbf{MXYNW}$ | [Norm] |
| 9 | $\mathbf{MXYNW} \leftrightarrow \mathbf{MXNW}$ | [Rep(6) |
| | $\mathbf{MXNYW} \leftrightarrow \mathbf{MXNW}$ | [Tr(7-9)] □ |

THEOREM 8. (Left Eliminate)

1 $\mathbf{X} \mapsto \mathbf{Y}$
$\overline{\mathbf{VYXZ} \leftrightarrow \mathbf{VXZ}}$

PROOF.

| | | |
|---|---|---|
| 2 | $\mathbf{X} \leftrightarrow \mathbf{YX}$ | [Suf(1)] |
| | $\mathbf{VYXZ} \mapsto \mathbf{VXZ}$ | [Rep(1,2)] □ |

THEOREM 9. (Drop)

1 $\mathbf{X} \mapsto \mathbf{VYZW}$
2 $\mathbf{X} \leftrightarrow \mathbf{V}$
$\overline{\mathbf{X} \mapsto \mathbf{VZ}}$

PROOF.

| | | |
|---|---|---|
| 3 | $\mathbf{VYZW} \mapsto \mathbf{XYZW}$ | [Rep(2)] |
| 4 | $\mathbf{X} \mapsto \mathbf{XYZW}$ | [Tran(1,3)] |
| 5 | $\mathbf{X} \mapsto \mathbf{XY}$ | [Dec(4)] |
| 6 | $\mathbf{XZW} \mapsto \mathbf{XY}$ | [Aug(5)] |
| 7 | $\mathbf{XZW} \leftrightarrow \mathbf{XYXZW}$ | [Suf(6)] |
| 8 | $\mathbf{XYXZW} \leftrightarrow \mathbf{XYZW}$ | [Norm] |
| 9 | $\mathbf{XZW} \leftrightarrow \mathbf{XYZW}$ | [Tran(7,8)] |
| 10 | $\mathbf{X} \mapsto \mathbf{XZW}$ | [Tran(4,9)] |
| 11 | $\mathbf{XZW} \mapsto \mathbf{VZW}$ | [Rep(2)] |
| 12 | $\mathbf{X} \mapsto \mathbf{VZW}$ | [Tran(10,11 |
| | $\mathbf{X} \mapsto \mathbf{VZ}$ | [Dec(12)] □ |

THEOREM 10. (Path)

| | |
|---|---|
| 1 | $\mathbf{X} \mapsto \mathbf{YW}$ |
| 2 | $\mathbf{Y} \leftrightarrow \mathbf{VMN}$ |
| | $\mathbf{X} \mapsto \mathbf{YMW}$ |

PROOF.

| | | |
|---|---|---|
| 3 | $\mathbf{X} \mapsto \mathbf{Y}$ | [Dec(1)] |
| 4 | $\mathbf{X} \mapsto \mathbf{VMN}$ | [Tran(2,3)] |
| 5 | $\mathbf{X} \mapsto \mathbf{YVMN}$ | [Union(3,4)] |
| 6 | $\mathbf{YVMN} \leftrightarrow \mathbf{YVMNM}$ | [Norm] |
| 7 | $\mathbf{X} \mapsto \mathbf{YVMNM}$ | [Trans(5,6)] |
| 8 | $\mathbf{X} \mapsto \mathbf{YM}$ | [Elim(2,7)] |
| 9 | $\mathbf{X} \mapsto \mathbf{YMYW}$ | [Union(1,8)] |
| 10 | $\mathbf{YMYW} \leftrightarrow \mathbf{YMW}$ | [Norm] |
| | $\mathbf{X} \mapsto \mathbf{YMW}$ | [Tr(1,10)] □ |

# 4. COMPLETENESS

In Section 4.1, we sketch the important elements of the proof for completeness of our OD axiomatization. We establish that ODs subsume FDs in Section 4.2, followed by the formal completeness proof of our axiomatization in Section 4.3.

## 4.1 Sketch of the Overall Proof

Our proof is constructive. To prove the axiomatization is complete, it suffices to demonstrate, for any set of ODs $\mathcal{M}$, a table **t** can be constructed that *satisfies* (Lemma 14), and is *complete* (Lemma 15) with respect to, $\mathcal{M}$ using $\mathcal{I}$, the axiomatization.

DEFINITION 11. (a table **t** satisfies $\mathcal{M}$) A table **t** *satisfies* $\mathcal{M}$ *iff* no OD that is derivable over $\mathcal{M}$ using $\mathcal{I}$ (thus, in $\mathcal{M}^+$) is falsified by the table **t**.

DEFINITION 12. (a table **t** is complete with respect to $\mathcal{M}$) A table **t** is *complete* with respect to $\mathcal{M}$ *iff* every OD that is constructible over the attributes that appear in $\mathcal{M}$ that is not derivable over $\mathcal{M}$ using $\mathcal{I}$ (thus, is not in $\mathcal{M}^+$) is falsified by the table **t**.

In Section 3.2 in Theorem 1, we proved the soundness of $\mathcal{I}$. Thus, any table that satisfies each OD in $\mathcal{M}$ satisfies $\mathcal{M}^+$, and no table that satisfies $\mathcal{M}$ can falsify any OD in $\mathcal{M}^+$. An OD $\mathbf{X} \mapsto \mathbf{Y}$ can be falsified in just two ways by a table. (See Theorem 15.) We name these two ways split and swap.

DEFINITION 13. (split) A split with respect to an OD $\mathbf{X} \mapsto \mathbf{XY}$ is a pair of tuples *t* and *s* in table **t**, such that $t_\mathbf{X} = s_\mathbf{X}$ but $(t_\mathbf{Y} \neq s_\mathbf{Y})$; that is, have the same value for **X** $(t_\mathbf{X} = s_\mathbf{X})$ but different values for **Y** $(t_\mathbf{Y} \neq s_\mathbf{Y})$. Thus, the split from **t** falsifies $\mathbf{X} \mapsto \mathbf{XY}$. (Consequently, $\mathbf{X} \mapsto \mathbf{Y}$ is falsified, too.) This just says that `set`(**X**) does not functionally determine `set`(**Y**).

DEFINITION 14. (swap) A swap with respect to an OD $\mathbf{XY} \leftrightarrow \mathbf{YX}$ is a pair of tuples *t* and *s* in table **t** such that $t_\mathbf{X} \prec s_\mathbf{X}$, but $s_\mathbf{Y} \prec t_\mathbf{Y}$; i.e., there exist tuples *t* and *s* in **t** such that $t_\mathbf{X} \prec s_\mathbf{X}$, but $s_\mathbf{Y} \prec t_\mathbf{Y}$; i.e., *t* comes before *s* in any stream satisfying `order by` **X**, but *s* comes before *t* in any stream satisfying `order by` **Y**. Thus, the swap from **t** falsifies $\mathbf{XY} \leftrightarrow \mathbf{YX}$. (Consequently, $\mathbf{X} \mapsto \mathbf{Y}$ is falsified, too.)

The table **t** that we construct for the set of order dependencies $\mathcal{M}$ will consist of two parts: split($\mathcal{M}$) and swap($\mathcal{M}$). We shall construct these two parts of **t** – the first half of the table, split($\mathcal{M}$), and the second half, swap($\mathcal{M}$) – in such a way that **t** satisfies $\mathcal{M}$. The purpose of split($\mathcal{M}$) will be to falsify every OD of the form $\mathbf{X} \mapsto \mathbf{XY}$ not in $\mathcal{M}^+$. The purpose of swap($\mathcal{M}$) will be to falsify every OD of the form $\mathbf{X} \mapsto \mathbf{Y}$, $\mathbf{XY} \leftrightarrow \mathbf{YX}$ not in $\mathcal{M}^+$ but for which $\mathbf{X} \mapsto \mathbf{XY}$ is in $\mathcal{M}^+$. (So $\mathbf{X} \mapsto \mathbf{Y}$ not in $\mathcal{M}^+$ by Theorem 15 appear) Thus, **t** is complete for $\mathcal{M}$.

DEFINITION 15. (split($\mathcal{M}$)) Split($\mathcal{M}$) is a table that demonstrates for each $\mathbf{X} \mapsto \mathbf{XY}$ which is not in $\mathcal{M}^+$ that $\mathbf{X} \mapsto \mathbf{XY}$ is falsified by split (and so, falsifies $\mathbf{X} \mapsto \mathbf{Y}$, too).

DEFINITION 16. (swap($\mathcal{M}$)) Swap($\mathcal{M}$) is a table that demonstrates for each $\mathbf{XY} \leftrightarrow \mathbf{YX}$ which is not in $\mathcal{M}^+$ that $\mathbf{XY} \leftrightarrow \mathbf{YX}$ is falsified by split (and so, falsifies $\mathbf{X} \mapsto \mathbf{Y}$, too).

Chain axiom is used to prove following two theorems.

THEOREM 11. (Partition)

| | |
|---|---|
| 1 | $\mathbf{X} \mapsto \mathbf{Y}$ |
| 2 | $\mathbf{X} \mapsto \mathbf{Z}$ |
| 3 | set(**Y**) = set(**Z**) |
| | $\mathbf{Y} \leftrightarrow \mathbf{Z}$ |

PROOF.

| | | |
|---|---|---|
| 4 | $\mathbf{X} \leftrightarrow \mathbf{YX}$ | [Suf(1)] |
| 5 | $\mathbf{X} \leftrightarrow \mathbf{X}$ | [Ref] |
| 6 | $\mathbf{X} \mapsto \mathbf{XY}$ | [Union(1,5)] |
| 7 | $\mathbf{X} \leftrightarrow \mathbf{XYX}$ | [Suf(6)] |
| 8 | $\mathbf{X} \leftrightarrow \mathbf{XY}$ | [Norm(7)] |
| 9 | $\mathbf{XY} \leftrightarrow \mathbf{YX}$ | [Tran(4,8)] |
| 10 | $\mathbf{XZ} \leftrightarrow \mathbf{ZX}$ | [2,4-9] |
| 11 | $\mathbf{X} \sim \mathbf{Y}$ | [(9)] |
| 12 | $\mathbf{X} \sim \mathbf{Z}$ | [(10)] |
| 13 | $\mathbf{XYZ} \leftrightarrow \mathbf{XYZ}$ | [Ref] |
| 14 | $\mathbf{XY} \leftrightarrow \mathbf{XZ}$ | [Elim(1,2,13)] |
| 15 | $\mathbf{Y} \sim \mathbf{Z}$ | [Chain(11-14)] |
| 16 | $\mathbf{YZ} \leftrightarrow \mathbf{ZY}$ | [(15)] |
| | $\mathbf{Y} \leftrightarrow \mathbf{Z}$ | [Norm(3,16)] □ |

THEOREM 12. (Downward Closure)

| | |
|---|---|
| 1 | $\mathbf{XY} \sim \mathbf{ZV}$ |
| | $\mathbf{X} \sim \mathbf{Z}$ |

PROOF.

| | | |
|---|---|---|
| 2 | $\mathbf{ZVXY} \mapsto \mathbf{Z}$ | [Ref] |
| 3 | $\mathbf{XYZV} \mapsto \mathbf{Z}$ | [Tran(1,2)] |
| 4 | $\mathbf{XYZV} \mapsto \mathbf{X}$ | [Ref] |
| 5 | $\mathbf{XYZV} \mapsto \mathbf{XZ}$ | [Union(3,4)] |
| 6 | $\mathbf{XYZV} \mapsto \mathbf{ZX}$ | [Union(3,4)] |
| | $\mathbf{X} \sim \mathbf{Z}$ | [Part(5,6)] □ |

In the table **t** that we construct, we shall use integer values for the cells. (A cell is a given column entry of a given row.) We construct table **t** by adding splits and swaps. We have to make sure that these pieces combined together do not interfere. That is why we formalize the notion of *append*. When we append two tables $\mathbf{t}_1$ and $\mathbf{t}_2$, we shall ensure that the resulting table cannot introduce any splits (except $\mathbf{X} \mapsto []$) or swaps beyond those that appear in $\mathbf{t}_1$ and in $\mathbf{t}_2$ alone (Lemma 9).

DEFINITION 17. (append) Appending two sub-tables $\mathbf{t}_1$ and $\mathbf{t}_2$ is accomplished by following steps:

1. Find the minimum value, x, over all cells of $\mathbf{t}_1$. Subtract x from all cells in $\mathbf{t}_1$. (Now its minimum value is zero.) Do the same for $\mathbf{t}_2$.
2. Find the maximum value, y, over all cells of $\mathbf{t}_1$. Add y + 1 to all cells in $\mathbf{t}_2$.The resulting table of the append is the union of $\mathbf{t}_1$ and $\mathbf{t}_2$ as adjusted in steps 1 and 2.

| A | B | C | D |
|---|---|---|---|
| 0 | 0 | 0 | 0 |
| 0 | 0 | 1 | 1 |

**Figure 4. Table $\mathbf{t}_1$.**

| A | B | C | D |
|---|---|---|---|
| 0 | 1 | 0 | 0 |
| 1 | 0 | 0 | 0 |

**Figure 5. Table $\mathbf{t}_2$.**

| A | B | C | D |
|---|---|---|---|
| 0 | 0 | 0 | 0 |
| 0 | 0 | 1 | 1 |
| 2 | 3 | 2 | 2 |
| 3 | 2 | 2 | 2 |

**Figure 6. $\mathbf{t}_1$ append $\mathbf{t}_2$.**

The table **t** we construct will be split($\mathcal{M}$) append swap($\mathcal{M}$) (which we call split-swap form). We shall construct split($\mathcal{M}$) in a way that is analogous to the construction in Ullman's proof of the completeness of Armstrong's axiomatization for FDs in [20]. This proves our axiomatization for ODs is sound and complete over FDs.

We shall construct swap($\mathcal{M}$) in a way to falsify each OD $\mathbf{X} \mapsto \mathbf{Y}$ not in $\mathcal{M}^+$ (but for which $\mathbf{X} \mapsto \mathbf{XY}$ is in $\mathcal{M}^+$). This construction will be more complex than for split($\mathcal{M}$). For each pair of attributes A and B from $\mathcal{M}$, we determine whether there needs to be a swap between A and B – a pair of tuples $s$ and $t$ such that $t_A \prec s_A$, but $s_B \prec t_B$ – and, if so, the *context* in which swap between A and B need to occur.

DEFINITION 18. (constant) An attribute A is called a *constant* with respect to $\mathcal{M}$ *iff* $[] \mapsto$ A is in $\mathcal{M}^+$. Call an attribute a *non-constant*, otherwise.

If an attribute is a constant, it means in any table that satisfies $\mathcal{M}$, it can have only a single value occurring in the table.

DEFINITION 19. (context) A set of non-constant attributes $\mathcal{X}$ with respect to $\mathcal{M}$ is a *context* of a swap *t*, *s* *iff* $t_{\mathcal{X}} = s_{\mathcal{X}}$. We say swap *t*, *s* is *in the context of* $\mathcal{X}$ iff $t_{\mathcal{X}} = s_{\mathcal{X}}$. (Note that a context for a swap *t*, *s* is not unique.)

By identifying the right contexts for swaps for each pair A and B, swap($\mathcal{M}$) will falsify each $\mathbf{X} \mapsto \mathbf{Y}$ not in $\mathcal{M}^+$ (but with $\mathbf{X} \mapsto \mathbf{XY}$ in $\mathcal{M}^+$), while not falsifying anything in $\mathcal{M}^+$ (Lemma 13). This step is the cornerstone of our proof for completeness.

Constructing table swap($\mathcal{M}$) is not straightforward. We are able to simplify the construction via structural induction. The hypothesis is as follows.

HYPOTHESIS 1 *(hypothesis). For some fixed integer K, for any set of ODs* $\mathcal{M}$ *composed over attributes* $\{E_1, \ldots, E_K\}$*, there exists a table* **t** *in split-swap form that satisfies, and is complete with respect to,* $\mathcal{M}$.

We prove the base case of this for $K \leq 2$ (in Lemma 11). We hypothesize this is true for any $\mathcal{M}$ with a fixed $K$ number of attributes. We then prove that for any $\mathcal{M}$ with $K+1$ attributes that the hypothesis remains true (Theorem 17). Proof of the induction hypothesis in essence completes the overall proof.

Induction provides us with a powerful mechanism within the proof. Consider any $\mathcal{M}$ with $K+1$ attributes. In the first case, if any of the attributes are constants with respect to $\mathcal{M}$, we can reduce the problem. We effectively *project out* those constant attributes from $\mathcal{M}$. This means we simply remove all occurrences of the attributes in the ODs. For example, if we are projecting out B and E, ABC $\mapsto$ DEF becomes AC $\mapsto$ DF. Call the result $\mathcal{M}$'. Then, $\mathcal{M}$' is over $K$ or fewer attributes. By the induction hypothesis there is a table **t**' which it satisfies, and is complete with respect to, $\mathcal{M}$'. We can show easily how to construct a table **t** from **t**' which must satisfy, and be complete with respect to, $\mathcal{M}$. This is established by Lemma 8.

LEMMA 8. *Let* **r** *be a table that satisfies, and is complete with respect to,* $\mathcal{M}$. *Let* Z *be an attribute not in* $\mathcal{M}$. *Construct table* **r**' *as* **r** *with an extra colum* Z, *and the same single value for* Z *in each row. Then* **r**' *satisfies, and is complete with respect to,* $\mathcal{M} \cup \{[] \mapsto \text{Z}\}$.

PROOF. It is straightforward that **r**' satisfies $\mathcal{M} \cup \{[] \mapsto \text{Z}\}$ because Z is a constant in **r**' and Z does not appear in $\mathcal{M}$. Clearly, **r**' falsifies each $\mathbf{X} \mapsto \mathbf{Y}$ that does not mention Z that **r** falsifies. For any $\mathbf{X} \mapsto \mathbf{Y}$ that mentions Z, it is equivalent to some OD that does not mention Z by the Replace rule, which has already been established. Thus **r**' satisfies, and is complete with respect to, $\mathcal{M} \cup \{[] \mapsto \text{Z}\}$. □

In the second case, we may assume $\mathcal{M}$ contains no constant attributes. When considering the pair A and B, if we find they require a swap in non-empty context $\mathcal{X}$, we can "freeze" the attributes of $\mathcal{X}$ to a single value. This is true, for any table that satisfies $\mathcal{M}' = \mathcal{M} \cup \{[] \mapsto \text{X}_1, \ldots, [] \mapsto \text{X}_n\}$, where $\mathcal{X} = \{\text{X}_1, \ldots, \text{X}_n\}$. Now, we have an instance with $K$ or fewer non-constants attributes. By our induction hypothesis, there exists a table $\mathbf{t}'$ in split-swap form that satisfies and is complete with respect to $\mathcal{M}'$. Note that $\mathcal{M}'^+ \supseteq \mathcal{M}^+$. Thus, $\mathbf{t}'$ does not falsify any ODs in $\mathcal{M}^+$. We append $\mathbf{t}'$ to the table **t** that we are constructing. (Appending these is safe, since $\mathcal{M}$ has no constants.) Our table swap($\mathcal{M}$) therefore is a recursive appending of (sub)tables.

There is the case of attributes A and B such that $\mathcal{M}$ dictates they must have a swap, but in the empty context {}. This time, we cannot use the induction hypothesis to construct the tuples for us ($\mathbf{t}'$), that do the job. For this case, however, we can construct two tuples directly that introduce a swap for A and B, but that do not introduce swaps between any other pair of attributes that would falsify any OD in $\mathcal{M}^+$. (The soundness of this step is established in Lemma 12.)

For the latter, we must show that, for each $\mathbf{X} \mapsto \mathbf{Y}$ not in $\mathcal{M}^+$ such that $\mathbf{X} \mapsto \mathbf{XY}$ is in $\mathcal{M}^+$, some sub-table in swap($\mathcal{M}$) by our construction does falsify it. This is done by proving there always is an attribute A in **X**, an attribute B in **Y**, and a swap between A and B in some context $\mathcal{W}$, which falsifies $\mathbf{X} \mapsto \mathbf{Y}$. (This is part of Lemma 15.) That completes the proof. These pieces are formally proved in the next two sections.

## 4.2 ODS Subsume FDS

In this section we show completeness of our axiomatization over FDs. This result is then used toward showing completeness over ODs.

THEOREM 13. (FD and OD correspondence) *For every instance* **r** *of relation* ***R***, $\mathcal{X} \to \mathcal{Y}$ *iff* $\mathbf{X} \mapsto \mathbf{XY}$*, for all lists* **X** *that order the attributes of* $\mathcal{X}$ *and all lists* **Y** *likewise for* $\mathcal{Y}$.

PROOF. **(IF)** If $\mathbf{X} \mapsto \mathbf{XY}$ holds by Lemma 1 $\mathcal{X} \to \mathcal{XY}$ is true. By Armstrong axiom, Reflexivity $\mathcal{XY} \to \mathcal{Y}$ holds. Therefore by Armstrong axiom, Transitivity $\mathcal{X} \to \mathcal{Y}$ is true.

**(ONLY IF)** If $\mathbf{X} \mapsto \mathbf{XY}$ does not hold, there exists *s*, $t \in \mathbf{r}$, such that $s_{\mathbf{X}} \preccurlyeq t_{\mathbf{X}}$ but $s_{\mathbf{XY}} \not\preccurlyeq t_{\mathbf{XY}}$. This implies that $s_{\mathbf{X}} = t_{\mathbf{X}}$ and $t_{\mathbf{Y}} \prec s_{\mathbf{Y}}$. Therefore $s_{\mathcal{Y}} \neq t_{\mathcal{Y}}$ and $s_{\mathcal{X}} = t_{\mathcal{X}}$ and $\mathcal{X} \to \mathcal{Y}$ is not true. □

THEOREM 14. (Permutation)

$$1 \quad \frac{\mathbf{X} \mapsto \mathbf{XY}}{\mathbf{X}' \mapsto \mathbf{X}'\mathbf{Y}'}$$

PROOF. Let $\mathbf{Y} = [\text{Y}_1, \text{Y}_2, \ldots, \text{Y}_n]$, $\forall k \in [1, n]$

| | | |
|---|---|---|
| 2 | $\mathbf{X} \mapsto \mathbf{X}\text{Y}_1 \ldots \text{Y}_k$ | [Dec(1)] |
| 3 | $\mathbf{X}'\mathbf{X} \mapsto \mathbf{X}'\mathbf{X}\text{Y}_1 \ldots \text{Y}_k$ | [Pref(2)] |
| 4 | $\mathbf{X}'\mathbf{X} \leftrightarrow \mathbf{X}'$ | [Norm] |
| 5 | $\mathbf{X}'\mathbf{X}\text{Y}_1 \ldots \text{Y}_k \leftrightarrow \mathbf{X}'\text{Y}_1 \ldots \text{Y}_k$ | [Norm] |
| 6 | $\mathbf{X}' \mapsto \mathbf{X}'\text{Y}_1 \ldots \text{Y}_k$ | [Tran(3-5)] |
| 7 | $\mathbf{X}' \leftrightarrow \mathbf{X}'$ | [Ref] |
| 8 | $\mathbf{X}' \mapsto \mathbf{X}'\text{Y}_k$ | [Drop(6,7)] |
| | $\mathbf{X}' \mapsto \mathbf{X}'\mathbf{Y}'$ | [Union(8)] □ |

An OD $\mathbf{X} \mapsto \mathbf{Y}$ can be falsified in two ways by a table (Theorem 15). That is why we introduced split and swap (Section 4).

THEOREM 15. (order dependency) $\mathbf{X} \mapsto \mathbf{Y}$ *holds iff* $\mathbf{X} \mapsto \mathbf{XY}$ *and* $\mathbf{XY} \leftrightarrow \mathbf{YX}$.

PROOF. **(IF)** If $\mathbf{X} \mapsto \mathbf{Y}$ holds then Suffix rule tells us, that $\mathbf{X} \leftrightarrow \mathbf{YX}$. $\mathbf{X} \mapsto \mathbf{X}$ follows from Reflexivity, therefore $\mathbf{X} \mapsto \mathbf{XY}$ by Union and $\mathbf{XY} \leftrightarrow \mathbf{YX}$ by Replace, Suffix and Normalization.

**(ONLY IF)** Suppose $\mathbf{XY} \leftrightarrow \mathbf{YX}$ and $\mathbf{X} \mapsto \mathbf{XY}$ hold. Hence, by Transitivity $\mathbf{X} \mapsto \mathbf{YX}$, which by Reflexivity and Transitivity tell us that $\mathbf{X} \mapsto \mathbf{Y}$. □

THEOREM 16. (ODs subsume FDs). *Given the set of ODs $\mathcal{M}$, OD axioms are sound and complete over functional dependencies.*

PROOF. Soundness is by Theorem 1, because of the correspondence between FDs and ODs (Theorem 13). The remaining step is to prove completeness over FDs, if $\mathcal{M} \models \mathcal{X} \rightarrow \mathcal{Y}$ then $\mathcal{M} \vdash_{\mathcal{I}} \mathcal{X} \rightarrow \mathcal{Y}$. This is equivalent to say if $\mathcal{M} \models \mathbf{X} \mapsto \mathbf{XY}$, *then* $\mathcal{M} \vdash_{\mathcal{I}} \mathbf{X} \mapsto \mathbf{XY}$ for all lists $\mathbf{X}$ that order the attributes of $\mathcal{X}$ and all lists $\mathbf{Y}$ likewise for $\mathcal{Y}$ by Theorem 13 and Permutation.

Firstly, we show that axioms for ODs imply Armstrong's axioms for FDs. We can do it because of soundness of axioms.

**$FD_1$ Reflexivity:** $\mathcal{Y} \subseteq \mathcal{X}$ implies $\mathcal{X} \rightarrow \mathcal{Y}$.

1. We are given that $\mathcal{Y}$ is a subset of $\mathcal{X}$.
2. Therefore, the normalization rule implies that an order dependency $\mathbf{X} \leftrightarrow \mathbf{XY}$ holds, for some list $\mathbf{X}$ that order the attributes of $\mathcal{X}$ and some list $\mathbf{Y}$ likewise for $\mathcal{Y}$.
3. Hence, Permutation and Theorem 13 implies that FD $\mathcal{X} \rightarrow \mathcal{Y}$ holds.

**$FD_2$ Augmentation:** $\mathcal{X} \rightarrow \mathcal{Y}$ implies $\mathcal{ZX} \rightarrow \mathcal{ZY}$.

1. Since we are given $\mathcal{X} \rightarrow \mathcal{Y}$, Theorem 13 tells us $\mathbf{X} \mapsto \mathbf{XY}$, for all lists $\mathbf{X}$ that order the attributes of $\mathcal{X}$ and all lists $\mathbf{Y}$ likewise for $\mathcal{Y}$.
2. By Reflexivity we can interfere $\mathbf{Z} \leftrightarrow \mathbf{Z}$, for all list $\mathbf{Z}$ that order the attributes of $\mathcal{Z}$. Hence, by Prefix rule $\mathbf{ZX} \mapsto \mathbf{ZXY}$ holds.
3. By Suffix $\mathbf{ZX} \leftrightarrow \mathbf{ZXYZX}$. $\mathbf{ZXYZX}$ may be normalized ($\mathbf{ZXYZX} \leftrightarrow \mathbf{ZXYZ}$).
4. By transitivity $\mathbf{ZX} \mapsto \mathbf{ZXYZ}$.Therefore by Permutation and Theorem 13 FD $\mathcal{ZX} \rightarrow \mathcal{ZY}$ holds.

**$FD_3$ Transitivity:** $\mathcal{X} \rightarrow \mathcal{Y}$ and $\mathcal{Y} \rightarrow \mathcal{Z}$ implies $\mathcal{X} \rightarrow \mathcal{Z}$.

1. We are given $\mathcal{X} \rightarrow \mathcal{Y}$, and $\mathcal{X} \rightarrow \mathcal{Z}$, so we may get $\mathbf{X} \mapsto \mathbf{XY}$ and $\mathbf{Y} \mapsto \mathbf{YZ}$ for all lists $\mathbf{X}$ that order the attributes of $\mathcal{X}$ and all lists $\mathbf{Y}$ likewise for $\mathcal{Y}$ and some list $\mathbf{Z}$ likewise for $\mathcal{Z}$ by Theorem 13.
2. It follows by Reflexivity that $\mathbf{X} \leftrightarrow \mathbf{X}$, so by Prefix rule we can infer that $\mathbf{XY} \mapsto \mathbf{XYZ}$.
3. Since $\mathbf{X} \leftrightarrow \mathbf{XY}$ follows by Suffix, Normalization and Transitivity, $\mathbf{X} \mapsto \mathbf{XYZ}$ follows from Transitivity.
4. Hence by Decomposition, Permutation and Theorem 13 FD $\mathcal{X} \rightarrow \mathcal{Z}$ is true.

However, this proves that axiom system comprising of inference rules $\mathcal{I}$ is sound and complete for the set of FDs $\mathcal{F}$. We would like to show it is true for set of ODs $\mathcal{M}$.

Let $\mathcal{M}' = \{\mathbf{X} \mapsto \mathbf{XY}, \mathbf{XY} \leftrightarrow \mathbf{YX} \mid \mathbf{X} \mapsto \mathbf{Y} \in \mathcal{M}\}$. Based on Theorem 15 $\mathcal{M}$ and $\mathcal{M}'$ are equivalent (Definition 9). Also let $\mathcal{F} = \{\mathcal{X} \rightarrow \mathcal{Y} \mid \mathbf{X} \mapsto \mathbf{XY} \in \mathcal{M}'\}$. Based on Permutation rule and Theorem 13 we know that any relation instance satisfying dependencies in $\mathcal{F}$ satisfies dependencies in $\mathcal{M}'$ and vice versa.

Let $\mathcal{X}^+$ [20], the closure of $\mathcal{X}$ (with respect to $\mathcal{F}$) be the set of attributes A such that $\mathcal{X} \rightarrow$ A can be deduced from $\mathcal{F}$ by Armstrong's axioms. We consider the relational instance $\mathbf{r}$ with the two rows shown in figure below.

| $\mathcal{X}^+$ attributes | | | | Other attributes | | | |
|---|---|---|---|---|---|---|---|
| 0 | 0 | … | 0 | 0 | 0 | … | 0 |
| 0 | 0 | … | 0 | 1 | 1 | … | 1 |

**Figure 7. A relation instance r showing that $\mathcal{M} \not\models \mathbf{X} \mapsto \mathbf{XY}$.**

Based on Ullman's [20] proof of soundness and completes of Armstrong's axioms, relation instance $\mathbf{r}$ shows that if $\mathcal{F}$ is the given set of dependencies, and $\mathcal{X} \rightarrow \mathcal{Y}$ cannot be proved by Armstrong, then $\mathbf{r}$ is a relation in which the dependencies of $\mathcal{F}$ hold but $\mathcal{X} \rightarrow \mathcal{Y}$ does not. That is, $\mathcal{F}$ does not logically imply $\mathcal{X} \rightarrow \mathcal{Y}$. This means the inference rules are sound and complete over $\mathcal{F}$. As there is no swaps in $\mathbf{r}$, we do not falsify anything in $\mathcal{M}'$, therefore $\mathcal{M}$, too. This ends the soundness and completeness proof for FDs over set of $\mathcal{M}$. □

## 4.3 Completeness of the OD Axiomatization

As discussed in Section 4 an OD can be falsified by a split or a swap. Using this, our proof for completeness is by case. If $\mathbf{X} \mapsto \mathbf{XY}$ is not in $\mathcal{M}^+$, there will be a split in the sub-table split($\mathcal{M}$) that we construct that falsifies $\mathbf{X} \mapsto \mathbf{XY}$, and so that falsifies $\mathbf{X} \mapsto \mathbf{Y}$ also. If $\mathbf{X} \mapsto \mathbf{Y}$ is not in $\mathcal{M}^+$, but $\mathbf{X} \mapsto \mathbf{XY}$ is, there will be a swap in sub-table swap($\mathcal{M}$) that falsifies $\mathbf{X} \mapsto \mathbf{Y}$.

LEMMA 9. *There is no split in $\mathbf{t}_1$ append $\mathbf{t}_2$ that is between rows from $\mathbf{t}_1$ and $\mathbf{t}_2$, respectively, besides* [] $\mapsto \mathbf{X}$ *for any* $\mathbf{X}$. *There is no swap in $\mathbf{t}_1$ append $\mathbf{t}_2$ that is between rows from $\mathbf{t}_1$ and $\mathbf{t}_2$, respectively.*

PROOF. Let *t* be a tuple in $\mathbf{t}_1$ and *s* be a tuple in $\mathbf{t}_2$. Since all values in *t* are less than all values in *s*, it is impossible for there to be a split (except [] $\mapsto \mathbf{X}$) or swap introduced between $\mathbf{t}_1$ and $\mathbf{t}_2$within $\mathbf{t}_1$ *append* $\mathbf{t}_2$ (Definition 17). □

We construct table $\mathbf{t}$ to satisfy, and to be complete with respect to, $\mathcal{M}$. Table $\mathbf{t}$ will be split($\mathcal{M}$) append swap($\mathcal{M}$), as introduced above. Note that by Theorem 15 these are the only two scenarios.

Table split($\mathcal{M}$) is constructed by appending two rows to the table, as in Figure 7 for each subset of attributes of $\mathcal{X}$ from $\mathcal{M}$.

LEMMA 10. (split($\mathcal{M}$) satisfies $\mathcal{M}$). *For any $\mathcal{M}$ with no constants, split($\mathcal{M}$) does not falsify any OD in $\mathcal{M}$.*

PROOF. The relational instance split($\mathcal{M}$) we have constructed contains splits, but no swaps. Therefore $\mathbf{X} \mapsto \mathbf{Y}$ could be only falsified by split. (Consequently, $\mathbf{X} \mapsto \mathbf{XY}$ is falsified, too.) But we know that we are sound and complete over set over FDs by Theorem 16 and by Lemma 9 appending of the tables does not introduce additional splits (except [] $\mapsto \mathbf{X}$) or swaps, therefore this is not possible. □

Table split($\mathcal{M}$) is based on table we constructed for $\mathcal{M}$ in the proof of Theorem 16, which establishes that ODs subsume FDs; that is, split($\mathcal{M}$) satisfies $\mathcal{M}$ and it is complete with respect to the OD of the form $\mathbf{X} \mapsto \mathbf{XY}$ – which are equivalent to FD statement (Theorem 13) – in that it falsifies each $\mathbf{X} \mapsto \mathbf{XY}$ not in $\mathcal{M}^+$ but which is composable over the attributes in $\mathcal{M}$. As constructed, split($\mathcal{M}$) introduces no *swaps*.

For swap($\mathcal{M}$) a natural approach would seem to be to construct the table incrementally, to falsify each OD not in $\mathcal{M}^+$, in turn, while ensuring we do not also falsify any OD in $\mathcal{M}^+$, in each step. This would be similar to how we constructed split($\mathcal{M}$). However, how to do this by a straightforward construction is not apparent. When considering how to falsify $\mathbf{X} \mapsto \mathbf{Y}$, which attributes from $\mathbf{X}$ and from $\mathbf{Y}$, respectively, should have a swap appear in the table? And how do we ensure that this swap does not falsify any OD in $\mathcal{M}^+$? Instead, we consider every pair of attributes, A and B, from the set of attributes in $\mathcal{M}$. We determine the relevant contexts, if any, in which a swap with respect to A and B must occur in swap($\mathcal{M}$).

The set(**XY**) is a context for A, B with respect to $\mathcal{M}$ *iff* **X**A ~ **Y** and **X** ~ **Y**B are in $\mathcal{M}^+$, but **X**A ~ **Y**B is not in $\mathcal{M}^+$. If there exists such a context for A, B, this indicates there should be a swap between A and B (to falsify **X**A ~ **Y**B). It also indicates the "context" of the swap, as the swap must not falsify **X**A ~ **Y** or **X** ~ **Y**B. One could imagine constructing a swap – a pair of rows *t* and *s* for this – by having $t_{\mathbf{XY}} = s_{\mathbf{XY}}$. That way, the swap *t*, *s* would not falsify **X**A ~ **Y** or **X** ~ **Y**B. But what should the values of *t* and *s* be outside of **XY**? We cannot construct *t* and *s* simply, ensuring the swap *s, t* does not falsify anything in $\mathcal{M}^+$. Instead, we use structural induction. Consider for now that **XY** is non-empty. If we added [] $\mapsto$ **XY** to $\mathcal{M}$ – call the result $\mathcal{M}'$ – **XY** can only have a single value in any table that satisfies $\mathcal{M}'$. Recall the hypothesis from Hypothesis 1in Section 4. We adopt this as our induction hypothesis. Assume our present $\mathcal{M}$ contains *K*+1 attributes. Then $\mathcal{M}'$ contains *K* or fewer attributes since [] $\mapsto$ **XY**. By our induction hypothesis, there is a table $\mathbf{t}'$ (see Figure 8) that satisfies, and is complete with respect to $\mathcal{M}'$. As **X**A ~ **Y**B is not in $\mathcal{M}^+$, it is not in $\mathcal{M}'^+$ either. Thus $\mathbf{t}'$ falsifies **X**A ~ **Y**B.

| Attributes of **XY** | | | | Other attributes | | | |
|---|---|---|---|---|---|---|---|
| 0 | 0 | … | 0 | $a_{1,1}$ | $a_{1,2}$ | … | $a_{1,i}$ |
| … | … | … | … | … | … | … | … |
| 0 | 0 | … | 0 | $a_{j,1}$ | $a_{j,2}$ | … | $a_{j,i}$ |

**Figure 8. A relation instance for K+1 non-constant attributes.**

Which context for A, B should we do this for? Not for all of them. It is the maximal contexts that are relevant. **X**, **Y** is a maximal context for A, B *iff* it is a context for A, B and there is no other context **X'**, **Y'** such that set(**X'Y'**) ⊃ set(**XY**).

Since we use induction in the proof, we need to prove a base case of the induction hypothesis. We prove it for the cases of $\mathcal{M}$ with 0, 1, and 2 non-constant attributes in the following Lemma.

LEMMA 11. (Induction base, $K \leq 2$). *For at most* $K \leq 2$ *attributes there exists a table **t** in split-swap form that satisfies and is complete with respect to* $\mathcal{M}$.

PROOF This can be directly shown by enumerating through all the possibilities. □

We have assumed so far that the (maximal) contexts, if any, for A, B are non-empty. There is the case where A, B has a single maximal context {}, the *empty* context. In this case, we cannot appeal to the induction hypothesis. Fortunately, such pair A, B will have special properties by virtue of the fact they have swapped orders only in the empty context. In fact, our sixth axiom schema speaks directly to this very case. (We likely would never have had the insight for the sixth axiom (schema) *Chain* had we not encountered this case while attempting to prove completeness.) In this case, we will be able to construct a two-row swap for A, B directly that does not falsify anything in $\mathcal{M}^+$.

LEMMA 12. (Empty context). *There exists a swap for A, B with the empty maximal context that satisfies* $\mathcal{M}$ *while falsifying A ~ B.*

PROOF. We construct a two-row swap with values 0 and 1 that falsifies A ~ B but cannot falsify anything in $\mathcal{M}^+$ as shown in Figure 9. For the latter, it suffices to prove that the swap does not falsify any C ~ D in $\mathcal{M}^+$. For A and B, they have opposite values in each row in the swap. For any C such that A ~ C is in $\mathcal{M}^+$, C must have the same value as A in each row. (Otherwise, A and C would have swapped values – 0 and 1 – between the two rows.) Likewise for B. And for any D such that C ~ D is in $\mathcal{M}^+$, D must have the same value as C (and so the same as A) in each row. And so forth. Of course, it would be impossible to construct our two rows if there is a chain connecting A and B through order-compatibility: A ~ $E_1$~ … ~ $E_n$ ~ B. If there were, we would need to set the value of each $E_1$~ … ~ $E_n$ the same as A's value *and* the same as B's value in each row. But A's and B's values differ. The Chain axiom schema (OD6) ensures there is no such chain from A to B. $E_i$A ~ $E_i$B is in $\mathcal{M}^+$, for each $E_i$, since the maximal context for A, B is []. If there were a chain A ~ $E_1$~ … ~ $E_n$ ~ B such that A ~ $E_1$ is in $\mathcal{M}^+$, $E_i$~ $E_{i+1}$ is in $\mathcal{M}^+$ for each *i* on $1,..,n-1$, and $E_n$ ~ B is in $\mathcal{M}^+$, then A ~ B is in $\mathcal{M}^+$ also, by the Chain axiom. Since we know that A ~ B is not in $\mathcal{M}^+$, there is no such Chain. Thus, our two rows are constructable. We can partition the attributes into three groups: those that must have the same values as A , those the same as B, and those for which it does no matter. Figure 9 shows the construction.

| A | B | A's group | | | B's group | | | Remaining attributes | | | |
|---|---|---|---|---|---|---|---|---|---|---|---|
| 0 | 1 | 0 | … | 0 | 1 | 1 | 1 | 0 | 0 | … | 0 |
| 1 | 0 | 1 | … | 1 | 0 | 0 | 0 | 1 | 1 | … | 1 |

**Figure 9. Swap for A, B with the empty maximal context.**

For attributes that do not match A or B, it is important we do not introduce swaps between them, as this could falsify something in $\mathcal{M}^+$. It suffices to use the same value for these in each row. Call the two-row swap in Figure 9 **r**. Table **r** satisfies $\mathcal{M}$. Assume otherwise: for **X**$\mapsto$**Y** ∈ $\mathcal{M}$, **r** falsifies it. Let **X** $\mapsto$ **Y** be over non-constants attributes, without loss of generality. Let E be the first element of **X**, and F of **Y**. If both E and F are from A, A's group or the remaining group attributes (as in Figure 9), or they are both from B or B's group attributes, then **X** and **Y** order the two tuples of **r** the same way. Therefore, E must be from one group, and F from the other. Since $\mapsto$**Y** ∈ $\mathcal{M}^+$, **X**~ **Y** ∈ $\mathcal{M}^+$ by Theorem 15. By the Downward Closure rule E~ F ∈ $\mathcal{M}^+$. Contradiction. □

Our proof obligation for swap($\mathcal{M}$), that it does not falsify any OD in $\mathcal{M}^+$ is proved in the following Lemma.

LEMMA 13. *(swap($\mathcal{M}$) satisfies $\mathcal{M}$). Assuming* Hypothesis 1, *for all* $\mathcal{M}$ *of K or fewer* non-constants *attributes, swap($\mathcal{M}$) does not falsify any OD in* $\mathcal{M}$.

PROOF. Hypothesis 1 is the key in proving that A, B do not falsify any OD in $\mathcal{M}^+$. When we consider pair A and B which requires a swap in non-empty context $\mathcal{X}$ we obtain $\mathcal{M}' = \mathcal{M} \cup \{[] \mapsto X_1, \ldots, [] \mapsto X_n\}$, where $\mathcal{X} = \{X_1, \ldots, X_n\}$. By our hypothesis, there exists a table $\mathbf{t}'$ in split-swap form that is satisfied and complete with respect to $\mathcal{M}'$. As $\mathcal{M}'^+ \supseteq \mathcal{M}^+$, therefore any ODs in $\mathcal{M}^+$ is not falsified.

None of the sub-tables falsifies any OD in $\mathcal{M}^+$, by the hypothesis in non-empty context and soundness of base cases (empty context and $K \leq 2$). As the table swap($\mathcal{M}$) is append-normalized, swap($\mathcal{M}$) does not falsify any OD in $\mathcal{M}^+$. □

LEMMA 14. (Satisfies). *Every OD that is derivable with respect to the axiomatization over* $\mathcal{M}$ *is not falsified by the table* **t**.

PROOF. The sub-tables split($\mathcal{M}$) and swap($\mathcal{M}$), as we construct them, are satisfied with respect to $\mathcal{M}$ (Lemma 10 and Lemma 13 respectively). If neither split($\mathcal{M}$) nor swap($\mathcal{M}$) falsifies any OD in $\mathcal{M}^+$, then **t** as split($\mathcal{M}$) append swap($\mathcal{M}$) cannot falsify any OD in $\mathcal{M}^+$ either (See Lemma 9). □

LEMMA 15. (complete). *Assuming* Hypothesis 1 *for all $\mathcal{M}$ constracted over K or fewer attributes, given any $\mathcal{M}$ constructed over K+1 attributes and none is a constant with respect to $\mathcal{M}$ (Definition* 18*), the table* **t** *= split($\mathcal{M}$) append swap($\mathcal{M}$) is complete with respect to $\mathcal{M}$.*

PROOF. Assume $\mathbf{X} \mapsto \mathbf{Y}$ over only non-constant attributes, is in the complement of $\mathcal{M}^+$ ($\mathbf{X} \mapsto \mathbf{Y} \notin \mathcal{M}^+$). Theorem 15 tells us that order dependency $\mathbf{X} \mapsto \mathbf{Y}$ holds iff $\mathbf{X} \mapsto \mathbf{XY}$ and $\mathbf{XY} \leftrightarrow \mathbf{YX}$.

**Case 1**.

$\mathbf{X} \mapsto \mathbf{Y} \notin \mathcal{M}^+$. We have already proven that for the scenario with $\mathbf{X} \mapsto \mathbf{XY}$ (FD) we are always complete (Theorem 16).

**Case 2**.

$\mathbf{X} \mapsto \mathbf{Y} \notin \mathcal{M}^+$, but $\mathbf{X} \mapsto \mathbf{XY} \in \mathcal{M}^+$. By Theorem 15 $\mathbf{X} \sim \mathbf{Y} \notin \mathcal{M}^+$. Find longest **P**A prefixing **X** such that:

1. $\mathbf{P} \sim \mathbf{Y} \in \mathcal{M}^+$
2. $\mathbf{P}\mathrm{A} \sim \mathbf{Y} \notin \mathcal{M}^+$

Find longest **Q**A prefixing **Y** such that:

3. $\mathbf{P}\mathrm{A} \sim \mathbf{Q} \in \mathcal{M}^+$
4. $\mathbf{P}\mathrm{A} \sim \mathbf{Q}\mathrm{B} \notin \mathcal{M}^+$
5. $\mathbf{P} \sim \mathbf{Q} \in \mathcal{M}^+$ [Downward Closure (1)]
6. $\mathbf{P} \sim \mathbf{Q}\mathrm{B} \in \mathcal{M}^+$ [Downward Closure (1)]
7. $\mathbf{P}\mathrm{A}\mathbf{Q}\mathrm{B} \leftrightarrow \mathbf{QP}\mathrm{AB} \in \mathcal{M}^+$ [Shift(3, [B ↔ B])]
8. $\mathbf{P}\mathrm{A}\mathbf{Q}\mathrm{B} \leftrightarrow \mathbf{PQ}\mathrm{AB} \in \mathcal{M}^+$ [Replace(5)]
9. $\mathbf{Q}\mathrm{B}\mathbf{P}\mathrm{A} \leftrightarrow \mathbf{PQ}\mathrm{BA} \in \mathcal{M}^+$ [Shift(6, [A ↔ A]]
10. $\mathbf{P}\mathrm{A}\mathbf{Q}\mathrm{B} \leftrightarrow \mathbf{Q}\mathrm{B}\mathbf{P}\mathrm{A} \notin \mathcal{M}^+$ [(4)]
11. $\mathbf{PQ}\mathrm{AB} \leftrightarrow \mathbf{PQ}\mathrm{BA} \notin \mathcal{M}^+$ [Transitivity(8,9,10]
12. $\mathbf{PQ}\mathrm{A} \sim \mathbf{PQ}\mathrm{B} \notin \mathcal{M}^+$ [11]

A and B have a swap within the context, $\mathcal{W}$ = set(**PQ**). In constructing swap($\mathcal{M}$), we considered all maximal contexts for A, B for which a swap is needed. Hence, we considered some superset $\mathcal{V} \supseteq \mathcal{W}$. If $\mathcal{V} \neq$ [], a sub-table that satisfies, and is complete with respect to $\mathcal{M} \cup \{[] \mapsto \mathrm{V}_1, \dots, [] \mapsto \mathrm{V}_n\}$, where $\mathcal{V} = \{\mathrm{V}_1, \dots, \mathrm{V}_n\}$ is appended in swap($\mathcal{M}$). This falsifies $\mathbf{W}\mathrm{A} \sim \mathbf{W}\mathrm{B}$, for all lists **W** that order the attributes of $\mathcal{W}$ (thus, falsifies $\mathbf{X} \mapsto \mathbf{Y}$). Else if $\mathcal{V}$ = [], we appended a swap *s, t* as in Figure 9 which falsifies A ~ B ([]A ~ []B). □

THEOREM 17. (soundness and completeness). *The set of the OD axioms $\mathfrak{I}$ ={OD1–OD6} is sound and complete.*

PROOF.

**Base case:** $\mathcal{M}$ with $K \leq 2$ attributes proved by Lemma 11. Assume Hypothesis 1 for all $\mathcal{M}$ composed over $K$ or fewer attributes.

**Induction step:** Consider an $\mathcal{M}$ over $K + 1$ attributes.

**Case 1**.

$\mathcal{M}$ contains constants attributes (Definition 18). Let $\mathcal{M}$' be $\mathcal{M}$ with these constants attributes removed. $\mathcal{M}$' has $K$ or fewer attributes. By the induction hypothesis (Hypothesis 1), there is **r**' which satisfies, and is complete with respect to, $\mathcal{M}$'. Lemma 8 guarantee we can construct **r** from **r**' that satisfies, and is complete with respect to, $\mathcal{M}$.

**Case 2.**

$\mathcal{M}$ contains no constants attributes. Lemma 15 establishes there exists an **r** that satisfies, an is complete with respect to, $\mathcal{M}$.□

# 5. RELATED WORK

Ordered sets and lattices have been a subject of research in mathematics [5]. In fact, our concept of OD is equivalent to *order-preserving mapping* between two ordered sets. The work in mathematics has concentrated on investigating properties of, and relationships between, ordered sets rather than among the mappings. To the best of our knowledge, no inference system for describing relationships between mappings has been proposed.

Order dependencies were introduced for the first time in the context of database systems in [7]. However, the type of orders, hence the dependencies defined over them, were different from the ones we presented here. A dependency $\mathcal{X} \rightsquigarrow \mathcal{Y}$ holds if order over the values of *each* attribute in $\mathcal{X}$ implies an order over the values of *each* attribute of $\mathcal{Y}$. (For simplicity, we use the arrow $\rightsquigarrow$ for different type of orders.) In other words, the dependency is defined over the sets of attributes rather than lists. The distinction between these two types of dependencies was later [13] aptly described as pointwise versus lexicographical order dependency. Formally, an instance satisfies a pointwise order dependency $\mathcal{X} \rightsquigarrow \mathcal{Y}$ if, for all tuples *s* and *t,* for every attribute A in $\mathcal{X}$, $s_{\mathrm{A}}$ *op* $t_{\mathrm{A}}$ implies that for every attribute B in $\mathcal{Y}$, $s_{\mathrm{B}}$ *op* $s_{\mathrm{B}}$, where *op* $\in \{<, =, >, \leq, \geq\}$. In [8] a sound and complete set of axioms for such dependencies is defined together with an analysis of the complexity of determining logical implication. An application of the dependencies for an improved index design is presented in [6].

Dependencies defined over lexicographically ordered domains were introduced in [13] under the name *lexicographically ordered functional dependencies*. Two other papers [14], [15] by the same author develop a theory behind both lexicographical as well as pointwise dependencies (the latter were somewhat simpler than the dependencies defined in [7]). A set of axioms (proved to be sound and complete) is introduced for pointwise dependencies, but – interestingly – not for lexicographical dependencies. Only a chase procedure is defined for the latter. An extension of relational algebra to ordered domains is presented in [15].

Sorting is at the heart of many database operations: sort-merge join, index generation, duplicate elimination, ordering the output through the `SQL` order-by operator, etc. The importance of sorted sets for query optimization and processing has been recognized very early on. Right from the start, the query optimizer of System R [16] paid particular attention to *interesting orders* by keeping track of all such ordered sets throughout the process of query optimization. In more recent research, [8] and [10] explored the use of sorted sets for executing nested queries. The importance of sorted sets has prompted the researchers to look *beyond* the sets that have been explicitly generated. Thus, [12] shows how to discover sorted sets created as generated columns via algebraic expressions. (In DB2, a generated column is a column that can be computed from other columns in the schema.)

For example, if column A is sorted, so is the generated column G defined as $\mathrm{G} = \mathrm{A}/100 + \mathrm{A} - 3$ (that is, A $\rightsquigarrow$ G). We show in [18] how to use relationships between sorted attributes discovered by reasoning over the physical schema. The axiomatization presented here provides a formal way of reasoning (discovering) previously unknown (or hidden) sorted sets. Based on this work, many other optimization techniques can also be adapted.

## 6. CONCLUSIONS AND FUTURE WORK

Ordering permeates databases, to such an extent that we take it for granted. It appears in many queries and is relatively expensive to perform. The goal of this paper was to develop a theory behind dependencies over lexicographically ordered sets. To the best of our knowledge, this is the first attempt at an axiomatization for such dependencies. We present that ODs subsumes FDs. We have also shown our inference rules for ODs are sound and complete.

Though now we conclude, the story of order dependencies is far from over. There is much more that can be done, and should be. Future work in this area should pursue two lines of research: on the one hand, further investigation of the theoretical questions; on the other hand, applications of the theoretical framework in a practical database setting. These are further things we plan to do in future work.

- One of the major practical applications which we are currently working on is a *theorem prover* [20]. Given a set of order dependencies $\mathcal{M}$ and an arbitrary dependency $\mathbf{X} \mapsto \mathbf{Y}$, we would like to *efficiently* decide whether $\mathcal{M}$ logically implies $\mathbf{X} \mapsto \mathbf{Y}$. Such a theorem prover would be a useful tool in query optimization.
- Integrity constraints have been widely used in query optimization through *query rewrites*. For example, functional dependencies have been shown to be useful in simplifying queries with `distinct`, `order by`, and `group by` operations [17], whereas inclusion dependencies can be used to remove certain joins over primary and foreign keys [4]. We believe that ODs can be used in similar ways to simplify queries with `order by` operation.
- We are exploring the use of ODs for *database design* [2]. FDs are by far the most common integrity constraints in the real world. The notion of the key derived from a given set of FDs is a fundamental to the relational model. The determination of ODs might be an important part of designing databases in the relational model, too. It can be used in database normalization and denormalization. Order dependencies can reveal redundancies that cannot be detected using functional dependencies alone. It would be an interesting research topic to extend the results obtained there to the design of relational databases.

## 7. ACKNOWLEDGMENTS

We thank Calisto Zuzarte and Wenbin Ma from IBM laboratory in Toronto for their encouragement and many helpful suggestions throughout the project.